\documentstyle[aps,prb,amsbsy,amssymb]{revtex}

\begin{document}
\twocolumn[\hsize\textwidth\columnwidth\hsize\csname
@twocolumnfalse\endcsname

\title{Columnar defects acting as passive internal field detectors}
\author{A.V. Silhanek$^1$\footnote{email:silhanek@cabbat1.cnea.gov.ar}, L. Civale$^1$, and M.A. Avila$^2$\footnote{present address: Ames Laboratory, Iowa State University}}
\address{$^1$Comisi\'{o}n Nacional de Energ\'{\i}a At\'{o}mica-Centro At\'{o}mico Bariloche and Instituto Balseiro, 8400 Bariloche, Argentina.\\
$^2$Instituto de F\'{i}sica ''Gleb Wataghin'', UNICAMP, 13083-970, Campinas - SP, Brazil.\\}

\date{\today}
\maketitle

\begin{abstract}

We have studied the angular dependence of the irreversible
magnetization of several YBa$_2$Cu$_3$O$_7$ and 2H-NbSe$_2$ single crystals with
columnar defects tilted off the c-axis. At high magnetic fields, the
irreversible magnetization $M_i(\Theta_H)$ exhibits a well known maximum
when the applied field is parallel to the tracks. As the field is
decreased below $H \sim 0.02 H_{c2}$, the peak shifts away from the tracks'
direction toward either the c-axis or the ab-planes. We demonstrate that
this shift results from the misalignment between the external and
internal field directions due to the competition between anisotropy and
geometry effects.

\end{abstract}

\vskip1pc] \narrowtext

\section{INTRODUCTION}

It is a well established fact that the presence of columnar
defects (CD) in high temperature superconductors (HTSC) enhance
the critical current ($J_c$) due to the strong pinning and the
inhibition of thermal wandering when the flux lines lay into these
tracks.\cite{civale91,konczykowski} The directional pinning
produced by these correlated structures becomes evident when the
angular dependence of $J_c$ is studied.\cite{evidence,interplay}
In the last decade several works have shown that a sharp peak in
$J_c(\Theta)$ appears when the applied field $\bf{H}$ is aligned
with the direction $(\Theta_D)$ of these linear
defects\cite{civale91,evidence,klein93,hardy96,zhukov97a,herbsommer98}
(Here $\Theta$ and $\Theta_D$ are the angles formed by the
crystallographic c-axis with $\bf{H}$ and the CD respectively).
However, this behavior only holds when $H$ is high enough to
ensure that the average vortex direction is parallel to $\bf{H}$.
At lower fields, both material anisotropy and geometry effects
become relevant and modify the vortex orientation,\cite{blatter94}
which consequently may not coincide with that of $\bf{H}$. Since
maximum pinning occurs when the vortex orientation (given by the
internal field $\bf{B}$ rather than by $\bf{H}$) is aligned with
the CD, $J_c(\Theta)$ should maximize at an angle $\Theta_{max}
\neq \Theta_D$. In other words, any misalignment between $\bf{B}$
and $\bf{H}$ manifests itself as a shift in the angular position
of the peak in $J_c(\Theta)$ with respect to the tracks direction.

In fact, we have recently shown in YBa$_2$Cu$_3$O$_7$(Y:123) that,
if the anisotropy effect dominates over the geometry effect (as
occurs in most cases for HTSC compounds, as we will show below)
for fields lower than $10 kOe$ the peak progressively departs from
the tracks orientation and shifts toward the c-axis\cite{evidence}
($\left|\Theta_{max}\right| < \left|\Theta_D\right|$). On the contrary, if the geometry
effect dominates over the anisotropy effect (as would be the case
in more isotropic materials), the maximum in $J_c(\Theta)$ should
move toward the ab-plane ($\left|\Theta_{max}\right| > \left|\Theta_D\right|$). Although the
latter effect has not been seen until now, Candia and
Civale\cite{candia} have experimentally demonstrated that in the
isotropic $Pb_{0.9}Tl_{0.1}$ alloy, the sample shape determines
the internal field direction. In that study it was shown that,
regardless of the direction of $\bf{H}$, at low fields the flux
lines remain almost locked to the sample normal because the system
gains energy by shortening the vortex length.

In this paper we present a study of the angular dependence of the
irreversible magnetization in several YBa$_2$Cu$_3$O$_7$ and
2H-NbSe$_2$ single crystals with a single set of CD tilted off the
c-axis. In particular, we study the misalignment between the
external and the internal field directions for several sample
aspect ratios and anisotropies. We use the uniaxial pinning of the
CD as a detector of the vortex orientation in the bulk of the
samples. We clearly demonstrate the influence of the crystal shape
in the determination of the vortex direction and find that the
angular behavior of the critical current is well described by the
competition between material anisotropy and sample geometry.

\section{THEORETICAL BACKGROUND}

A complete description of the vortex lattice behavior in a superconducting material should include geometry effects, mass anisotropy, vortex-vortex and vortex-defects interactions.
Whenever the system is in thermodynamic equilibrium, the internal field ${\bf B}$ is determined by minimization of the free energy\cite{blatter94} \mbox{$G({\bf B})=F({\bf B})-\frac{B^2}{8\pi}+\frac{({\bf B}-{\bf H}){\bf M}}{2}$}. In this expression the magnetization ${\bf M}$ and the applied field ${\bf H}$ are related by \mbox{${\bf H}={\bf B}-4\pi(1-\hat\nu){\bf M}$}, where $\hat\nu$ is the tensor of demagnetization factor. The components of $\hat\nu$ at the sample principal axes are $(\nu_x,\nu_y,\nu_z)$, with $\nu_x+\nu_y+\nu_z=1$. We adopt the notation that $z$ coincides with the crystallographic c-axis, and that the $x$ axis is perpendicular to both c and $\bf H$. Standard minimization of $G({\bf B})$ with respect to $B_y$ and $B_z$ gives,

\begin{equation}
\frac{\partial F}{\partial B_i}=\frac{B_i}{4\pi}-\frac{B_i-H_i}{4\pi \left(1-\nu_i\right)},~~~~ where~~~ i=y,z.
\label{eq:derivatives}
\end{equation}

In the intermediate fields regime ($H_{c1} \ll H
\ll H_{c2}$) the free energy $F({\bf B})$ is given by\cite{blatter94},

\begin{equation}
F = \frac{B^2}{8 \pi} + \frac{\Phi_0 B \epsilon_{\Theta_B}}{2(4 \pi
\lambda_{ab})^2} ln \left[ \frac{H^c_{c2}}{\epsilon_{\Theta_B} B} \right],
\label{eq:F}
\end{equation}
where ${\Theta_B}$ is the direction of the internal field,
$\epsilon_{\Theta_B}=\sqrt{\cos^2\Theta_B+\epsilon^2\sin^2\Theta_B}$, the anisotropy $\epsilon = m_{ab}/m_c <1$, and $\lambda_{ab}$ is the penetration depth for ${\bf H}\parallel$c.

By replacing (\ref{eq:F}) into (\ref{eq:derivatives}) we obtain

\begin{equation}
sin(\Theta_B-\Theta)=-\frac{f(\nu_y,\nu_z,\epsilon)\sin(2\Theta_B)}{8\kappa^2}
\frac{\ln h+1}{h}\label{eq:scaling}
\end{equation}
where $f(\nu_y,\nu_z,\epsilon)=(1-\nu_z)-(1-\nu_y)\epsilon^2$ and the reduced field $h= H/H_{c2}(\Theta_B,T)$.

The result (\ref{eq:scaling}) only assumes uniaxial anisotropy and
the coincidence of one principal axis with the c-axis, and it
shows that under those very general conditions the misalignments
due to both mass anisotropy and sample geometry have the same
field and temperature dependence. The function
$f(\hat\nu,\epsilon)$, which contains the combined effects of
geometry and anisotropy, is the key ingredient of the low field
behavior, as its sign determines whether $\Theta_B$ leads or lags
behind $\Theta$.

To be more specific, let's consider the typical platelike shape of all the single crystals of both Y:123 and NbSe$_2$ used in this study, with thickness $t$ along the c-axis much smaller than the lateral dimensions $L_x$ and $L_y$. To a first approximation $\nu_x=t/L_x$ and $\nu_y=t/L_y$, thus \mbox{$\nu_x, \nu_y, (1-\nu_z)~\ll~1$}. If the material is strongly anisotropic and the crystal is not too thin, then \mbox{$(1-\nu_z)~>~(1-\nu_y)\epsilon^2$}, thus $f~>~0$ and \mbox{$\Theta_B~>~\Theta$}. We will call this the ``anisotropy-dominated'' situation. In contrast, for thin enough samples of a not too anisotropic material \mbox{$(1-\nu_z)~<~(1-\nu_y)\epsilon^2$}, so \mbox{$\Theta_B~<~\Theta$}. This is what we will call the ``geometry-dominated'' case. The extreme limit of this case, with an infinite slab ($\nu_x=\nu_y=0$) and ignoring the anisotropy, has been discussed by Klein et al.\cite{klein93}.
It is also worth to note that for an infinite cilinder with axis perpendicular to ${\bf H}$, where the geometry effects are expected to cancel out, $\nu_x=0$ and \mbox{$\nu_y=\nu_z=\frac {1}{2}$}, thus \mbox{$f \propto \left( 1-\epsilon^2 \right)$} and  eq.(\ref{eq:scaling}) reduces to the well known expression for the bulk.\cite{evidence,blatter94}

Eq.(\ref{eq:scaling}) allows us to determine which should be the
vortex direction $\Theta_B$ for a given angle $\Theta$ of the
controlled variable ${\bf H}$. To check whether this model
describes the basic vortex lattice behavior when $\bf{H}$ is
tilted away from the c-axis, we will use the CD as internal field
detectors taking profit from the fact that $J_c$ maximizes when
$\Theta_B=\Theta_D$. Thus, if we know $\Theta_D$ and
$\Theta_{max}$ (the angular position of the maximum in $J_c$), we
are able to determine the misalignment
$\Theta_B-\Theta=\Theta_D-\Theta_{max}$. Although the {\it sign}
of such misalignment is solely determined by the sign of $f$, its
{\it magnitude} also depends on additional factors such as
$\sin(2\Theta_B)$ and $\kappa^2$. Besides that, the misalignment
is strongly temperature and field dependent. It is easy to see
from eq. (\ref{eq:scaling}) that $\Theta_B \rightarrow \Theta$ for
large enough $h$. Throughout this paper we will change each of
these factors in order to show that eq. (\ref{eq:scaling})
satisfactorily accounts for the observed properties.

At this point it is important to note that, although the misalignment between $\bf{B}$ and $\bf{H}$ is a low field effect, eq.~(\ref{eq:scaling}) can only be used in the field range where eq. (\ref{eq:F}) is valid, i.e., for $H \gg H_{c1}$. It turns out that all our data are well described by eq. (\ref{eq:scaling}). However, the very dilute vortex limit is conceptually interesting, and for completeness we will discuss it in the last section.

\section{EXPERIMENTAL DETAILS}

The Y:123 single crystals used in this work were grown by the
self-flux method as described in ref.[\cite{paco94}] and exhibit a
critical temperature of $T_c=92 K$. We also performed measurements
on single crystals of the layered superconductor NbSe$_2$ with
$T_c=7.2 K$. In all cases, columnar defects off the c-axis were
introduced by irradiation with $300 Mev$ $Au^{26+}$ ions, using
the TANDAR accelerator facility (Buenos Aires, Argentina). Table I
summarizes the information about geometrical dimensions, mass
anisotropy, dose-equivalent matching field $B_\Phi$ and angle
$\Theta_D$ of the CD with respect to the c-axis, for all the
crystals measured.

The dc magnetization measurements were performed in a commercial
SQUID magnetometer with two pickup coils, and both components
(longitudinal $M_l$ and transverse $M_t$) were recorded. Samples
could be rotated around the axis perpendicular to both the c-axis
and the CD. Two measuring procedures were used. The first one
consisted of collecting a set of isothermal magnetization loops,
each one recorded at a fixed angle $\Theta$. In the second one,
that we developed more recently,\cite{avila01} the sample is
rotated in steps at fixed T and H, and ${\bf M}$ is recorded at a
dense set of orientations (the sample is not rotating during the
measuring scan). By appropriate proccessing, both methods allows
us to obtain the irreversible magnetization vector ${\bf
M_i}(H,T,\Theta)$, as described
previously\cite{evidence,candia,avila01,anomalous}.

In general, the relation (given by the critical state Bean model)
between ${\bf M_i}$ and the various components of the anisotropic
$J_c$ is complicated. However, for thin platelike samples, and as
long as $\tan\Theta < L_y/t \sim \nu_y^{-1}$ (which in all the
crystals used in our study is true for almost all field
orientations, except in a very narrow angular range near $\Theta =
90^{\circ}$), due to purely geometrical constrains the following
conditions are satisfied:\cite{zhukov-geometry} First, the
pinning-related persistent currents flow essentially parallel to
the ab-planes, thus only one component of $J_c$ is involved.
Second, ${\bf M_i}$ is almost normal to the sample surface. Third,
the geometrical factor that relates the modulus $M_i$ with $J_c$
is almost independent of $\Theta$. As a consequence, a measure of
$M_i(\Theta,H,T)$ is equivalent to a measure of the in-plane
$J_c(\Theta,H,T)$.

\section{EXPERIMENTAL RESULTS}

\subsection{anisotropy-dominated case}

Figure 1 shows the angular dependence of the irreversible magnetization
for Y:123 crystals $A$ and $B$ at $T=70K$ for several fields. The most
evident feature in this figure is the asymmetry  of $M_i(\Theta)$
around $\Theta=0$ arising in the uniaxial pinning due to the CD. For sample
$A$ at $H>10 kOe$, we observe the well known peak at the tracks angular
position $\Theta_D=32^{\circ}$, and at lower fields the peak
progressively shifts away from $\Theta_D$ toward the c-axis. A more
complete set of curves showing the shift for this crystal at various T and H
can be found in ref.\cite{evidence}. A similar behavior is observed in
crystal $B$, although the shift turns out to be smaller than in $A$.
These two crystals have the same anisotropy and irradiation conditions,
but different shapes. Thus, at the same T and H all factors in eq.~(\ref{eq:scaling}) are identical, except for $f(\hat\nu,\epsilon)$. As
seen in table~I, the difference in demagnetizing factors results in a
smaller $f(\hat\nu,\epsilon)$ for sample $B$ than for $A$. Hence, the
misalignment in sample $B$ is expected to be smaller, as indeed
observed.

\subsection{compensated case}

A striking result predicted by eq.~(\ref{eq:scaling}) is that the
competing effects (anisotropy and geometry) could be exactly compensated if
one were able to tune the demagnetizing factors and the anisotropy in
order to get $f(\hat\nu,\epsilon)=0$, a condition that is satisfied
for $1-\nu_z=\left(1-\nu_y\right)\epsilon^2$. For
the Y:123 single crystals used in this work, with $\epsilon \sim 1/7$,
this requires extremely thin samples with a big area. Table I shows that
crystals $C$ and $D$ almost exactly satisfy this compensating
condition, as the absolute values of their $f(\hat\nu,\epsilon)$ are,
respectively, a factor of $\sim 20$ and $\sim 10$ smaller than in $A$. Fig.~2
shows the angular dependence of $M_i$ for these two crystals at
$T=60K$. For the sake of clarity only a reduced set of fields is shown, and
some curves have been shifted vertically. Since the CD orientations
$\Theta_D$ in samples $C$ and $D$ are different, in order to compare them
we set the abscise as the relative angle $\Theta-\Theta_D$. In this
figure we clearly observe that the peaks remain locked at the tracks
direction even for the lowest fields, in complete agreement with the
expectation. The same behavior was observed for other temperatures.

\subsection{geometry-dominated case}

So far, the four samples studied were Y:123 crystals with the same anisotropy but
different geometries. On these samples we observed that the peak either
shifts in the direction of the c-axis or almost does not deviate from
the CD direction. As pointed out previously, this behavior arises from
the strong anisotropy effect in this material. In order to change the
sign of the deviation (i.e., a shift toward the ab-plane), we  need to
reduce the anisotropy effects. (Table I shows that crystal $D$ has \mbox{$f<0$},
thus strictly speaking it is in the geometry-dominated case, but the
shift is too small to be detected).

To that end we decided to measure NbSe$_2$ single crystals, which
have $\epsilon \sim 1/3$, making the anisotropy effect about 5
times smaller than in Y:123. Besides this, very large and thin
NbSe$_2$ crystals can be readily found, and they can be easily cut
to obtain the desired shape. So we irradiated a rectangular
crystal (labelled as sample $E$), such that it is in the extreme
geometry-dominated case, with $f(\hat\nu,\epsilon) \ll 0$.
Fig.~3(a) shows $M_i(\Theta)$ for this sample at $T=4.4 K$ for
several $H$. At high fields we observe a peak at the tracks'
direction $\Theta_D=27^{\circ}$. As field decreases the peak
becomes broader and, in contrast to the Y:123 observed behavior,
it progressively moves away from $\Theta_D$ toward the ab-plane,
in agreement with a negative $f(\hat\nu,\epsilon)$.

The conclusive evidence that the function $f(\hat\nu,\epsilon)$
dominates the behavior of the misalignment $\Theta_B-\Theta$ comes from
samples $F$ and $G$, which are pieces of crystal $E$. These samples were
obtained by cutting the sample $E$ along a line parallel to its
shortest side, in such a way that the demagnetizing factor $\nu_x$
remains unaltered, but $\nu_y$
increases. In this way, the absolute value of the function
$f(\hat\nu,\epsilon)$ was progressively reduced, i.e., we moved away from the
``extreme geometry-dominated case'' and approached the ``compensated case''
(see Table I). According to eq.~(\ref{eq:scaling}), the deviation of the
maximum in $M_i(\Theta)$ for given $H$ and $T$ should become progressively
smaller for crystals $F$ and $G$ as compared to crystal $E$. This is in
fact observed, as demonstrated in Fig.~4, where $\Theta_{max}-\Theta_D$ for crystals $E$, $F$ and $G$ is plotted as a function of $h=H/H_{c2}(T,\Theta)$.

The misalignments for all the Y:123 crystals shown in Figs.~(1)
and (2) are also included in Fig.~4. Thus, this figure summarizes
all the samples studied in the present work, at various
temperatures and fields. The three possible low field behaviors
are clearly visible: anisotropy-dominated (upward curvature),
geometry-dominated (downward curvature) and compensated (almost
horizontal curves). It is worth to note that, in all the
non-compensated cases and for both materials, the misalignment
between $\bf{B}$ and $\bf{H}$ becomes relevant for fields below a
certain characteristic field $H \sim 0.02 H_{c2}$.

\subsection{Quantitative test of the model}

The above results clearly demonstrate that the qualitative differences in the low field behavior are controlled by the factor $f(\hat\nu,\epsilon)$. We now want to verify whether the $H$ and $T$ dependence of the shift is well described by the model. According to eq.~(\ref{eq:scaling}) these two variables appear only through the combination $h= H/H_{c2}(\Theta_B,T)$. Thus, $\left| \Theta_D-\Theta_{max} \right|$ should increase not only with decreasing $H$ at fixed $T$, as already seen in figs.~1 and 3(a), but also with decreasing $T$ at fixed $H$, due to the increase in $H_{c2}(T)$. This second expectation is also verified, as shown in Fig.~3(b)
where $M_i(\Theta)$ for sample $E$ was plotted at constant field $H=0.15 kOe$ for several temperatures.

The equivalence between the variations in T and H is quantitatively verified in the main panel of figure~5, where $\sin(\Theta_{max}-\Theta_D)/f(\hat\nu,\epsilon)$ is plotted as a function of $h$ for the two sets of data shown in Fig.~3(a) and (b). We observe quite a good scaling, thus confirming that $h$ is the appropriate variable. The upper critical field values $H_{c2}(kG)=51.8\left(1-t\right)$ were taken from the literature,\cite{detrey} thus the
superposition of the two curves involves no free parameters.

Finally, we analize the quantitative effect of the factor $f(\hat\nu,\epsilon)$. This factor is a constant for a given sample, so it is the same for all the data in the main panel of fig. 5. In contrast, in the inset we show the same scaling procedure for the crystals $E, F$ and $G$ at $T=4.4K$, so now $f(\hat\nu,\epsilon)$ is different for each sample, while all the other parameters remain identical. We again obtain a good superposition of the data, although the scaling is poorer than in the main panel, probably due to the damage produced in the crystal after each cut process.

The solid line in the main panel of Fig. 5 depicts the expectation of eq. (\ref{eq:scaling}), with $\Theta_B = \Theta_D = 27^\circ$ as experimentally determined from the location of the maximum at high fields, and a single fitting parameter $\kappa = 5.6$. The same curve is shown in the inset. Although the value $\kappa = 5.6$ is smaller than the accepted value\cite{detrey} $\kappa \sim 9$, a similar discrepancy was reported by Zhukov et al. when studying the lock-in effect in Y:123 by both twin boundaries\cite{zhukov97a} and columnar defects.\cite{zhukov98a} Moreover, in previous studies\cite{evidence,avila01} in Y:123 and Er:123 with CD, we had also found that the results are well described using values of the penetration depth $\lambda$ smaller than the accepted ones. We will discuss this issue in the next section.

\section{THE INFLUENCE OF THE CD}

It is important to keep in mind that the shift in $\Theta_{max}$
at low $h$ is {\it not} due to the CD. We are only using them as a
passive tool to measure the vortex direction in the bulk of the
samples, which is not easy to do by other methods. In fact, the
pinning of the CD is not contained in eq.(\ref{eq:scaling}), which
arises from the minimization of a free energy, and thus describes
a state of thermodynamic equilibrium. It is obvious, on the other
hand, that the uniaxial pinning of these correlated structures is
relevant and should be included in the  analysis. This is usually
done\cite{hardy96,blatter94} by adding to the free energy
(\ref{eq:F}) a term $F_{pin}$ that accounts for the correction to
the vortex self-energy due to the CD, and then comparing the
energy of alternative configurations.

This additional contribution depends on the orientation of the vortices, $F_{pin}=F_{pin}\left(\Theta_B\right)$, and it is always negative, reflecting the fact that, for $\Theta_B \neq \Theta_D$, the CD promote the formation of staircase vortices whose self energy is lower than that of a straight vortex at the same average orientation. $F_{pin}\left(\Theta_B\right)$ decreases as $\Theta_B$ approaches $\Theta_D$ due to the increase of the core trapped fraction, and it minimizes for that orientation, when the vortex cores are totally trapped into the tracks.\cite{blatter94} The key point in the context of the present study, however, is that the incorporation of $F_{pin}\left(\Theta_B\right)$ into the scenario does not modify the previous results, as we show below.

Let's first consider that $\bf {H}$ is applied at the angle
$\Theta = \Theta_{max}$ such that, in the absence of pinning and
according to eq.~(\ref{eq:scaling}), the vortices would be at the
angle $\Theta_B = \Theta_D$. If we now ``turn on''
$F_{pin}\left(\Theta_B\right)$, the only effect will be to deepen
the already existing minimum of the free energy at this
orientation, without changing the angle.

Let's now suppose that $\bf {H}$ is applied at an angle $\Theta$
slightly smaller or slightly larger than $\Theta_{max}$. In the
absence of pinning vortices would respectively orient at angles
$\Theta_B$ slightly smaller or slightly larger than $\Theta_D$,
according to eq.~(\ref{eq:scaling}). The addition of the term
$F_{pin}\left(\Theta_B\right)$ will now shift the vortices towards
$\Theta_D$, that is, a kind of effective {\it angular attractive
potencial} towards the CD orientation will develop. In particular,
for $\left|\Theta-\Theta_{max}\right| < \varphi_L$, the influence
of $F_{pin}\left(\Theta_B\right)$ will be so strong that the
system will minimize its free energy by orienting the vortices
exactly along the CD. This is the well known lock-in effect,
namely that the internal field remains locked to $\Theta_D$ over a
finite range of $\Theta$, and $\varphi_L(h)$ is called the {\it
lock-in angle}. In previous works\cite{evidence,avila01} we have
extensively studied this effect, that manifests in our
measurements as a {\it plateau} in $M_i(\Theta)$ of width
$2\varphi_L$. Note that the center of the plateau coincides with
$\Theta_{max}$. Thus, although the relation $\Theta_B$ vs $\Theta$
will be modified by the CD, the angle $\Theta_{max}$,
experimentally defined as the maximum in $M_i(\Theta)$ or as the
center of the plateau where necessary, will still be given by
eq.~(\ref{eq:scaling}).

As a final comment it is worth to mention that, although
$\varphi_L\propto 1/h$, similarly to the misalignment between $\bf
{B}$ and $\bf {H}$, the physical origin of both effects is totally
different. Indeed, in NbSe$_2$ we observe the shift but not the
plateau associated to the lock-in, probably because the pinning of
the CD is less effective in this material as compared to the Y:123
and Er:123 compounds.

\section{THE VERY LOW FIELD LIMIT}
For completeness, we now consider the very low field limit, although it does not apply to our data. In this limit only nearest neighbours (NN) vortex-vortex interactions need to be taken into account, thus

\begin{equation}
F \approx \frac{B H_{c1}\left(\Theta_B\right)}{4\pi}\left[1+g\left({\bf B}\right)\right]
\label{eq:Fdilute}
\end{equation}
where $H_{c1}\left(\Theta_B\right) \approx H_{c1}^c\epsilon_{\Theta_B}$.

In an {\it isotropic} material, the sum of the 6 equal contributions from the NN gives $g\left({\bf B}\right)=k r^{-1/2}e^{-r}$, where $k=\frac{6}{\ln\kappa}\sqrt{\frac{\pi}{2}} \sim 1$; $r=a/\lambda$ and
$a=\sqrt{\frac{\Phi_0}{B}}$ is the vortex lattice parameter. In this limit $r \gg 1$, thus $g\left({\bf B}\right) \ll 1$. In the {\it anisotropic} case, $g\left({\bf B}\right)$ must be modified to account for the angular dependence of $\lambda$, and for the distortion of the triangular lattice. Regardless of the details, it is clear that in this case $g\left({\bf B}\right)$ will still be exponentially small at low enough vortex densities, and the same is true for the derivatives $\left(\partial g/\partial B_i\right)$. Thus, replacement of (\ref{eq:Fdilute}) into eq. (\ref{eq:derivatives}) gives

\begin{equation}
H_{c1}^c\frac{\epsilon^2 \sin\Theta_B}{\epsilon_{\Theta_B}}
\left[1+\eta_y\left(\bf B\right)\right]=\frac{H_y-\nu_yB_y}{1-\nu_y}
\label{eq:Bydilute}
\end{equation}

\begin{equation}
H_{c1}^c\frac{\cos\Theta_B}{\epsilon_{\Theta_B}}
\left[1+\eta_z\left(\bf B\right)\right]=\frac{H_z-\nu_zB_z}{1-\nu_z}
\label{eq:Bzdilute}
\end{equation}
where $\eta_i\left(\bf B\right)$, which account for NN interactions, are exponentially small in the limit of $B \rightarrow 0$.

By taking the ratio of both components, we can calculate the relation between $\Theta$ and $\Theta_B$ for the first vortex to penetrate,

\begin{equation}
\epsilon^2\tan\Theta_B \approx \frac{\left(H_y-\nu_yB_y\right)\left(1-\nu_z\right)}
{\left(H_z-\nu_zB_z\right)\left(1-\nu_y\right)}
\approx \frac{\left(1-\nu_z\right)}{\left(1-\nu_y\right)}\tan \Theta
\label{eq:thetaBdilute}
\end{equation}

So, in the very dilute limit the system is in the
geometry-dominated case for
\mbox{$\epsilon^2\left(1-\nu_y\right)>\left(1-\nu_z\right)$} and
in the anisotropy-dominated case for
\mbox{$\epsilon^2\left(1-\nu_y\right)<\left(1-\nu_z\right)$},
exactly the same result that we had found for intermediate fields.

Eq.~(\ref{eq:scaling}) predicts that the misalignment grows indefinitely as $h \rightarrow 0$, even allowing for $\Theta_B$ and $\Theta$ to lay in different quadrants. As an extreme case, that equation has no real solution when the absolute value of its RHS becomes larger than unity. These clearly unphysical results are just a manifestation of the inapplicability of eq.~(\ref{eq:scaling}) at very low fields. As an example, we can estimate the field at which the peak in $M_i(\Theta)$ is predicted to coincide with the ab-plane ($\Theta=90^{\circ}$) for sample $E$. The result is $h \approx 0.0013$, or $H/H_{c1} \approx 0.09$, [that is $H \approx 25Oe$ in fig ~3(a)], clearly below the lower limit of validity. (Note that, due to the strong demagnetizing effects, vortex penetration becomes energetically favourable not at $H_{c1}$ but rather at a much smaller field $H_{c1}^* \sim \left(1-\nu_z\right)H_{c1} \sim 0.02H_{c1}$ for sample $E$.) In contrast, eq.~(\ref{eq:thetaBdilute}) indicates that for this sample at very low fields the peak should be observed at $\Theta \approx 88^{\circ}$, as marked with an arrow in fig.~3(a).

Finally, we must note that the influence of the uniaxial pinning of the CD turns progressively stronger as $h$ decreases and the term $F_{pin}$ becomes a significant fraction of the total free energy. At low enough fields the lock-in angle $\varphi_L\propto 1/h$ covers most of the angular range (this is essentially what Klein et al.\cite{klein93} called {\it flux-flop} in their early work) thus producing a very broad plateau that turns unapplicable our method to determine $\Theta_{max}$.

\section{CONCLUSION}

In summary, we performed a detailed study of the influence of anisotropy and sample geometry  in the determination of the vortex orientation in bulk superconductors. We showed that these effects are relevant at low fields and become negligible at high fields. On top of that, we developed a model that correctly accounts for the sample shape, mass anisotropy, temperature and field dependencies of the misalignment between the applied field and the vortex direction. We demonstrated that the sign of the misalignment is solely determined by a function that contains the combined effects of the anisotropy and the demagnetizing factors, and we presented experimental examples of the three possible situations, namely anisotropy-dominated, geometry-dominated and compensated cases. We finally discussed the very low vortex density limit.

An important practical consequence of these results is that studies of the pinning
properties of tilted CD that are based solely on measurements at
$\bf H \parallel$CD, or on comparison of this orientation with a
few others, give valid information at high fields, but are
misleading at low fields: vortices are just not oriented in the
right direction. To avoid this problem, a rather complete
knowledge of the angular dependent response is required.

\section{ACKNOWLEDGMENTS}
We are pleased to thank David J. Bishop for providing us with the NbSe$_2$ single crystals and Juan A. Herbsommer for his valuable help in selecting crystals for irradiation.
This work was partially supported by ANPCyT, Argentina, PICT 97
No. 01120 and by FAPESP, Brazil, Procs. \#96/01052-7
and \#96/05800-8. A.V.S. would like to thank the CONICET for
financial support.

\section{REFERENCES}

\bibliographystyle{prsty}

\pagebreak

\onecolumn
\begin{table}[ht]
\centering \caption{Irradiation and shape specifications for all the
crystals studied.}
\begin{tabular}[b]{l c c c c c c c c c c}
Crystal & material & $\epsilon^{-1}$ & $B_{\Phi}(kOe)$ &
$\Theta_D$ & $t(\mu m)$ & $L_y(\mu m)$ & $L_x(\mu m)$ & $\nu_y(\times 10^{-3})$ & $\nu_x(\times 10^{-3})$ & $f(\hat\nu,\epsilon) (\times 10^{-3})$ \\
\hline
A & $YBa_2Cu_3O_7$ & 7 & 30 & $32^{\circ}$ & 8.5 & 210 & 630 & 40 & 13.5 & +34 \\

B & $YBa_2Cu_3O_7$ & 7 & 30 & $32^{\circ}$ & 20.9 & 715 & 2150 & 29 & 9.7 & +19 \\

C & $YBa_2Cu_3O_7$ & 7 & 57 & $30^{\circ}$ & 11.5 & 1050 & 1050 & 11 & 11 & +1.8 \\

D & $YBa_2Cu_3O_7$ & 7 & 22 & $57^{\circ}$ & 4.3 & 381 & 762 & 11.3 & 5.6 & -3.2 \\

E & $NbSe_2$ & 3 & 0.5 & $27^{\circ}$ & 7.7 & 765 & 640 & 10.1 & 12 & -67 \\

F & $NbSe_2$ & 3 & 0.5 & $27^{\circ}$ & 7.7 & 585 & 640 & 13.2 & 12 & -63.5 \\

G & $NbSe_2$ & 3 & 0.5 & $27^{\circ}$ & 7.7 & 419 & 640 & 18.4 & 12 & -58 \\

\end{tabular}
\end{table}

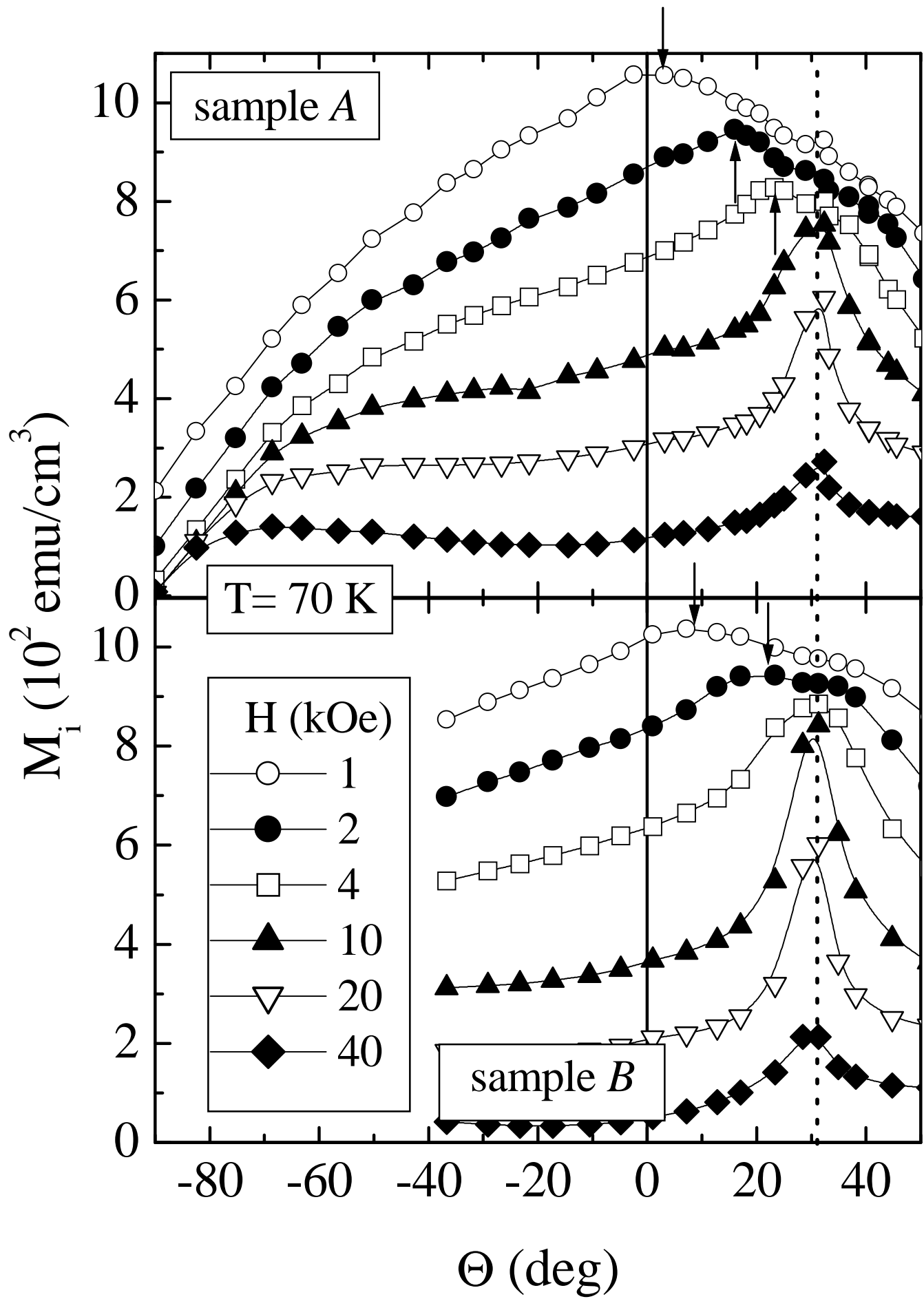
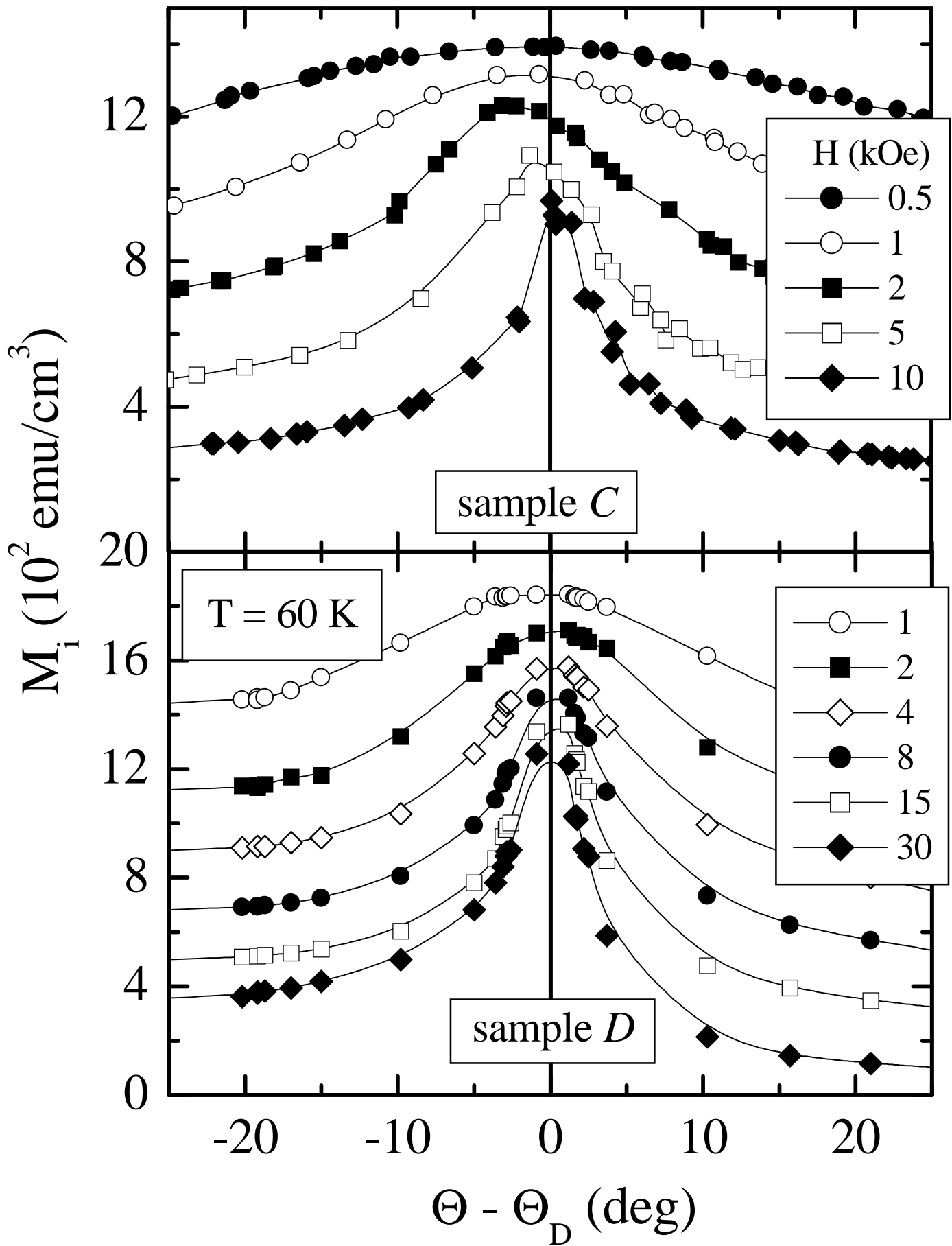
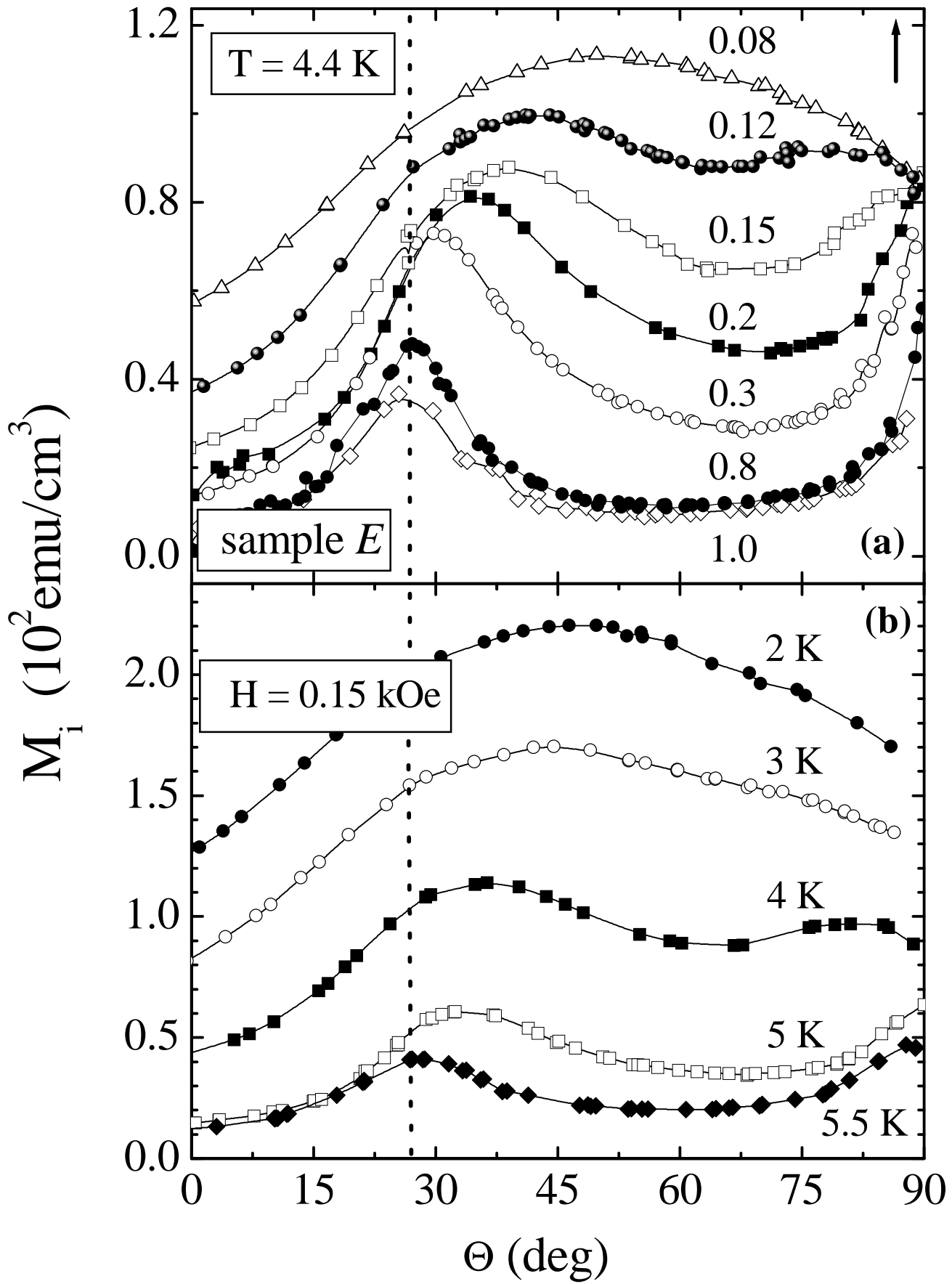
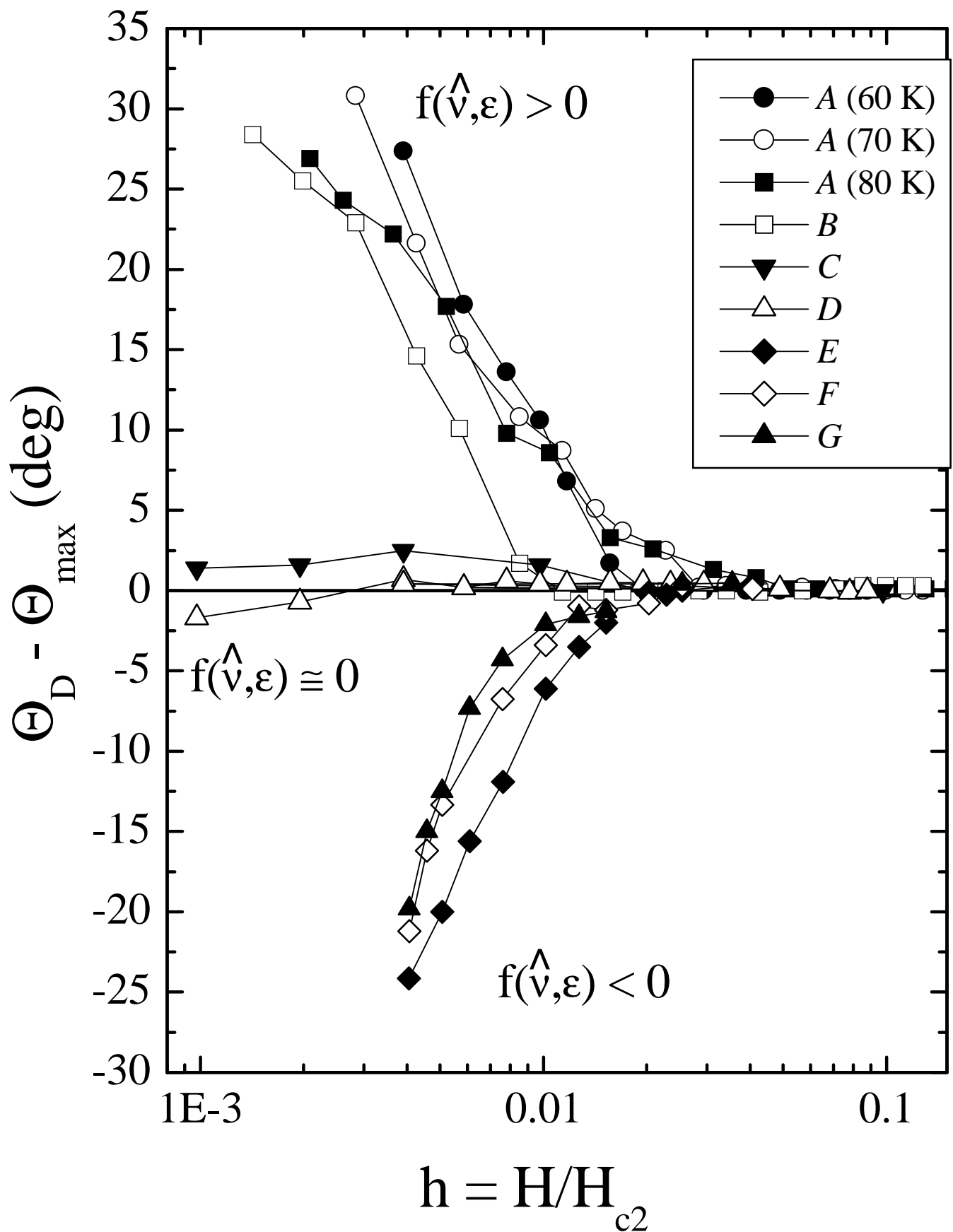
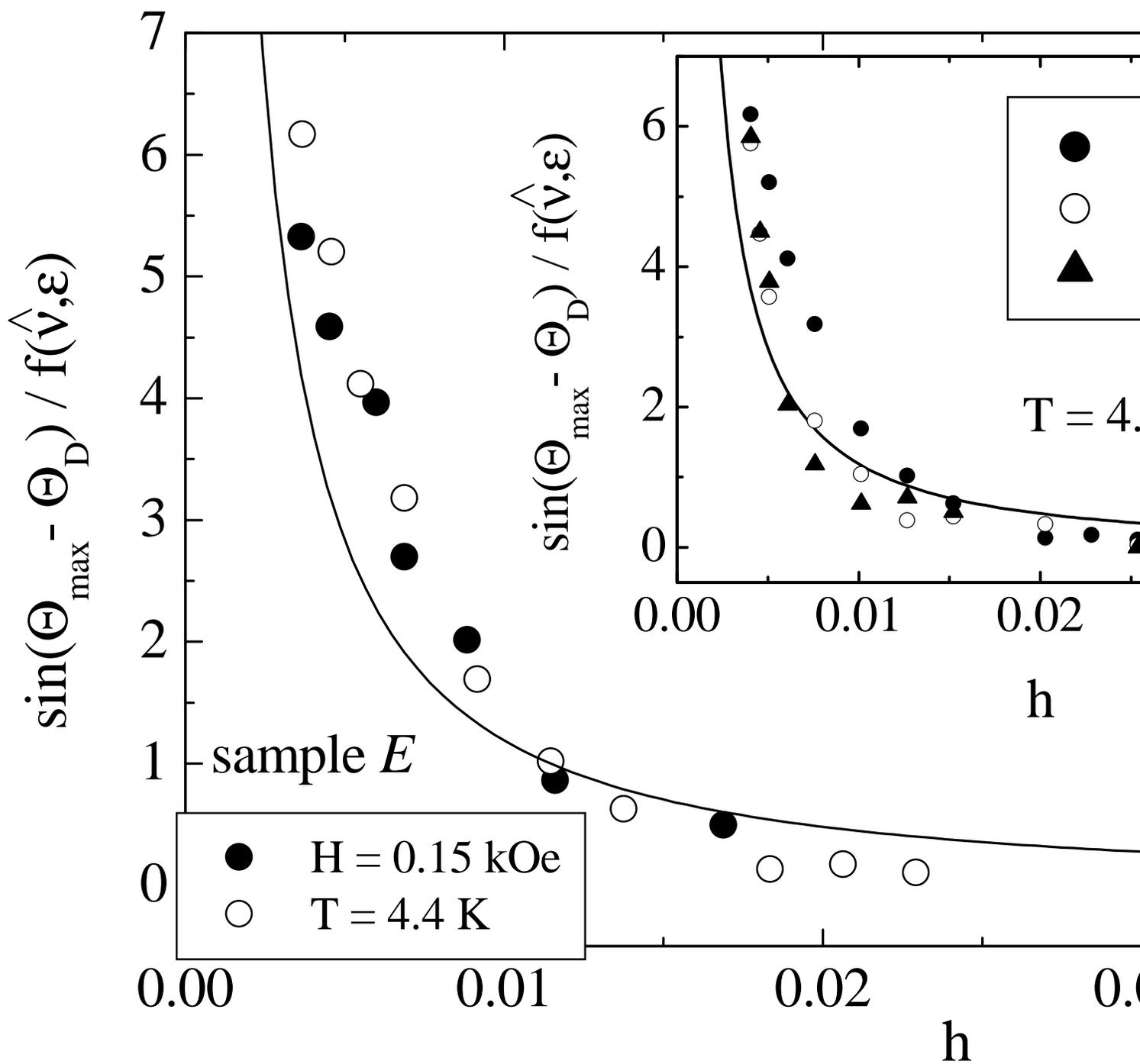
\end{document}